\documentstyle[12pt]{article}
\newread\epsffilein    
\newif\ifepsffileok    
\newif\ifepsfbbfound   
\newif\ifepsfverbose   
\newdimen\epsfxsize    
\newdimen\epsfysize    
\newdimen\epsftsize    
\newdimen\epsfrsize    
\newdimen\epsftmp      
\newdimen\pspoints     
\def\epsfbox#1{\global\def\epsfllx{72}\global\def\epsflly{72}%
   \global\def\epsfurx{540}\global\def\epsfury{720}%
   \def\lbracket{[}\def\testit{#1}\ifx\testit\lbracket
   \let\next=\epsfgetlitbb\else\let\next=\epsfnormal\fi\next{#1}}%
\def\epsfgetlitbb#1#2 #3 #4 #5]#6{\epsfgrab #2 #3 #4 #5 .\\%
   \epsfsetgraph{#6}}%
\def\epsfnormal#1{\epsfgetbb{#1}\epsfsetgraph{#1}}%
\def\epsfgetbb#1{%
\openin\epsffilein=#1
\ifeof\epsffilein\errmessage{I couldn't open #1, will ignore it}\else
   {\epsffileoktrue \chardef\other=12
    \def\do##1{\catcode`##1=\other}\dospecials \catcode`\ =10
    \loop
       \read\epsffilein to \epsffileline
       \ifeof\epsffilein\epsffileokfalse\else
          \expandafter\epsfaux\epsffileline:. \\%
       \fi
   \ifepsffileok\repeat
   \ifepsfbbfound\else
    \ifepsfverbose\message{No bounding box comment in #1; using defaults}\fi\fi
   }\closein\epsffilein\fi}%
\def\epsfsetgraph#1{%
   \epsfrsize=\epsfury\pspoints
   \advance\epsfrsize by-\epsflly\pspoints
   \epsftsize=\epsfurx\pspoints
   \advance\epsftsize by-\epsfllx\pspoints
   \epsfxsize\epsfsize\epsftsize\epsfrsize
   \ifnum\epsfxsize=0 \ifnum\epsfysize=0
      \epsfxsize=\epsftsize \epsfysize=\epsfrsize
     \else\epsftmp=\epsftsize \divide\epsftmp\epsfrsize
       \epsfxsize=\epsfysize \multiply\epsfxsize\epsftmp
       \multiply\epsftmp\epsfrsize \advance\epsftsize-\epsftmp
       \epsftmp=\epsfysize
       \loop \advance\epsftsize\epsftsize \divide\epsftmp 2
       \ifnum\epsftmp>0
          \ifnum\epsftsize<\epsfrsize\else
             \advance\epsftsize-\epsfrsize \advance\epsfxsize\epsftmp \fi
       \repeat
     \fi
   \else\epsftmp=\epsfrsize \divide\epsftmp\epsftsize
     \epsfysize=\epsfxsize \multiply\epsfysize\epsftmp
     \multiply\epsftmp\epsftsize \advance\epsfrsize-\epsftmp
     \epsftmp=\epsfxsize
     \loop \advance\epsfrsize\epsfrsize \divide\epsftmp 2
     \ifnum\epsftmp>0
        \ifnum\epsfrsize<\epsftsize\else
           \advance\epsfrsize-\epsftsize \advance\epsfysize\epsftmp \fi
     \repeat
   \fi
   \ifepsfverbose\message{#1: width=\the\epsfxsize, height=\the\epsfysize}\fi
   \epsftmp=10\epsfxsize \divide\epsftmp\pspoints
   \vbox to\epsfysize{\vfil\hbox to\epsfxsize{%
      \includegraphics{#1}%
      \hfil}}%
\epsfxsize=0pt\epsfysize=0pt}%

{\catcode`\%=12 \global\let\epsfpercent=
\long\def\epsfaux#1#2:#3\\{\ifx#1\epsfpercent
   \def\testit{#2}\ifx\testit\epsfbblit
      \epsfgrab #3 . . . \\%
      \epsffileokfalse
      \global\epsfbbfoundtrue
   \fi\else\ifx#1\par\else\epsffileokfalse\fi\fi}%
\def\epsfgrab #1 #2 #3 #4 #5\\{%
   \global\def\epsfllx{#1}\ifx\epsfllx\empty
      \epsfgrab #2 #3 #4 #5 .\\\else
   \global\def\epsflly{#2}%
   \global\def\epsfurx{#3}\global\def\epsfury{#4}\fi}%
\def\epsfsize#1#2{\epsfxsize}
\textwidth=16 cm
\textheight=625 pt
\oddsidemargin=0 pt
\voffset=-1 cm

\begin{document}
\newcommand{\bfk}{{\mbox{\boldmath $k$}}}
\newcommand{\bfp}{{\mbox{\boldmath $p$}}}
\newcommand{\bfB}{{\mbox{\boldmath $B$}}}
\newcommand{\bfsig}{{\mbox{\boldmath $\sigma$}}}
\newcommand{\bfn}{{\mbox{\boldmath $n$}}}
\newcommand{\bfe}{{\mbox{\boldmath $e$}}}
\newcommand{\bfq}{{\mbox{\boldmath $q$}}}
\newcommand{\one}{{\mbox{1\hskip-0.5mm l}}}
\newcommand{\be}{\begin{eqnarray}} \newcommand{\ee}{\end{eqnarray}}
\thispagestyle{empty}
\noindent
hep-ph/9603333 \hfill
PITHA 96/07 \rightline{March, 1996} \rightline{ }
\renewcommand{\thefootnote}{\fnsymbol{footnote}}
\begin{center}  {\Large\bf Spin-Spin Correlations of Top Quark Pairs
    at Hadron Colliders}
 \end{center}
 \vskip 10 true mm
\begin{center} {Arnd Brandenburg\\
    \vskip 3mm
    \it Institut f\"ur Theoretische Physik, Physikzentrum\\
    Rheinisch-Westf\"alische Technische Hochschule Aachen\\
    52056 Aachen, Germany} \end{center} \vskip 10 true mm\bigskip
\centerline{\large\bf Abstract} \smallskip
\noindent  Top quark pairs are produced 
with strongly correlated spins in the partonic reactions $q\bar{q}\to
t\bar{t}$ and $gg\to t\bar{t}$. A complete description of these
effects  in terms of the spin density matrix of the $t\bar{t}$ system
in leading order QCD is given.
We further discuss the prospects to observe
the spin-spin correlations at $p\bar{p}$ and $pp$ colliders by measuring
suitable angular correlations among the $t$ and $\bar{t}$ decay
products.  \vfil\eject 
\section{Introduction}

After the recent discovery of the top quark \cite{CDFD0}, the detailed
study of the properties of this particle will be a major subject of
experiments at the (upgraded) Tevatron and at future colliders.  
An intriguing feature of
the top quark is that due to its heaviness it decays on average before
it can form hadronic bound states.  Moreover, the typical spin flip
time is much larger than the lifetime of the top \cite{Bigi}. Thus a
possible polarization of the $t\bar{t}$ system induced by the
production mechanism will be transferred to its decay products. Due to
the dominant parity violating decay mode $t\to Wb$ the $t$ and
$\bar{t}$ ``self-analyze'' their spins. The spin information may be
extracted by forming angular correlations among the $t$ and $\bar{t}$
decay products, thus allowing for a variety of tests of the standard
model (SM) and extensions thereof (see, e.g. \cite{spinreffirst}-
\cite{spinreflast}).\par While at hadron colliders the 
{\it{longitudinal}} polarization of top quarks is practically zero due
to parity invariance of QCD, a nonzero polarization {\it{transverse}}
to the production plane is induced by absorptive parts at the one-loop
level. The prospects to observe this order $\alpha_s$ effect at the
Tevatron and the Large Hadron Collider (LHC) have been studied in
detail in \cite{BBU} (see also \cite{Goldstein1},\cite{KLY}).  Apart
from this single quark polarization, the $t$ and $\bar{t}$ are
produced with {\it{strongly correlated spins}} in $q\bar{q}\to
t\bar{t}$ and $gg\to t\bar{t}$, which are the dominant partonic
production processes at the Tevatron and the LHC, respectively.  In
fact, these spin-spin correlations are of order one at the level of
the partonic reactions.  It is the aim of this letter to discuss the
prospects to unravel these effects. An experimental verification of
the feasibility to extract spin information on top quarks would
clearly be important for any further proposals to study quantities
related to the top spin. \par In the next section we will give a
general description of the $t\bar{t}$ spin state in terms of a spin
density matrix.  QCD induced spin-spin correlations may be described
in general by four ``structure functions''. The leading order results
for these functions are given. Observables built from the spin
operators of $t$ and $\bar{t}$ allow to discuss the magnitude of the
correlations at parton level.  In Section 3 we will construct
observables which are directly measurable (on an event by
event basis) in $p\bar{p},pp\to t\bar{t}X$ with subsequent $t\bar{t}$
decays.

\section{The spin density matrices for $q\bar{q}\to t\bar{t}$ and
 $gg\to t\bar{t}$}
Although the spin of an unstable particle produced in high-energy
reactions is no directly observable quantity, it is very useful to
introduce the concept of a {\it{spin density matrix}} for the
$t\bar{t}$ system.  Using the narrow width approximation for the top
quark, we may view the reactions considered here as the production and
subsequent decay of on-shell top quark pairs. The spin information may
then be extracted on a statistical basis from the decay products of
the $t$ and $\bar{t}$.

\par
We first discuss the reaction $q(p_1)+\bar{q}(p_2)\to
t(k_1)+\bar{t}(k_2)$, where the momenta refer to the partonic c.m.
system, $\bfp_1+\bfp_2=0$.  The complete spin information is encoded
in the (unnormalized) spin density matrix $R^q$,
\begin{eqnarray} \label{density1}
R^q_{\alpha_1\alpha_2,\beta_1\beta_2}(\bfp,\bfk)=
\frac{1}{4}\frac{1}{N_C^2}\sum_{{\mbox{\scriptsize{colors}}},
q\bar{q}{\mbox{\scriptsize{\ spins}}}}
\langle t(k_1,\alpha_1)\bar{t}(k_2,\beta_1)|{\cal{T}}|
q(p_1),\bar{q}(p_2)\rangle^* & &\nonumber\\
\langle t(k_1,\alpha_2)
\bar{t}(k_2,\beta_2)|{\cal{T}}|
q(p_1),\bar{q}(p_2)\rangle. & &
\end{eqnarray}
Here, $\alpha,\ \beta$ are spin indices, $N_C$ denotes the number of
colors, $\bfp=\bfp_1,\ \bfk=\bfk_1$ and the sum runs over the colors
of all quarks and over the spins of $q$ and $\bar{q}$. The factor
$1/4\cdot 1/N_C^2$ takes care of the averaging over spins and colors
in the initial state. The matrix structure of $R^q$ in the spin spaces
of $t$ and $\bar{t}$ is
\begin{eqnarray}\label{density2}
R^q=A^q\one\otimes \one+ \bfB^q_t\cdot\bfsig\otimes \one+\bfB^q_{\bar{t}}\cdot
\one\otimes\bfsig+C^q_{ij}\sigma^i\otimes\sigma^j,
\end{eqnarray}
where $\sigma^i$ are the Pauli matrices and the first (second) factor
in the tensor products refers to the $t$ ($\bar{t}$) spin space. (The
spin operators of $t$ and $\bar{t}$ are simply given by
$\bfsig/2\otimes \one$ and $\one\otimes\bfsig/2$.) For a detailed
discussion of $R^q$ and its symmetry properties, see \cite{Bernbra}.
We will now specialize on reactions mediated by strong interactions.
Imposing P and CP invariance on $R_q$ and using rotational invariance,
we are left with the following structures:
\begin{eqnarray}
A^q \ &=&\ \frac{8\pi \hat{s}}{\beta}\frac{d\hat{\sigma}^q}{dz},\nonumber \\ 
\bfB^q_t \ &=&\  \bfB^q_{\bar{t}}\ =\ b_3^q(\hat{s},z)\frac{{\bfp\times \bfk}}
{|\bfp\times \bfk|}\\
C_{ij}^q \ &=&\ c_0^q(\hat{s},z)\delta_{ij}+ 
c_4^q(\hat{s},z)\hat{p}_i\hat{p}_j+c_5^q(\hat{s},z)\hat{k}_i\hat{k}_j
+c_6^q(\hat{s},z)(\hat{p}_i\hat{k}_j+\hat{k}_i\hat{p}_j).
\end{eqnarray} 
Here, $\beta=\sqrt{1-4m_t^2/\hat{s}}$ is the velocity of the top quark
in the partonic c.m. system, $\hat{\bfp}=\bfp/|\bfp|,\ 
\hat{\bfk}=\bfk/|\bfk|$, $z=\hat{\bfp}\cdot\hat{\bfk}$ is the
scattering angle and $\hat{s}=(p_1+p_2)^2$ is the partonic c.m. energy
squared. (The notation, in particular the numbering, is adopted from
\cite{Bernbra}.)  For $gg\to t\bar{t}$ we have an analogous
decomposition with $1/N_C^2 \to 1/(N_C^2-1)^2$ in (\ref{density1}).
The functions $b_3^i,\ i=q,g$ derive from absorptive parts in the
scattering amplitude \cite{BBU}. Here we are interested in the
spin-spin correlation functions $c_0^i,\ c_4^i,\ c_5^i$ and $c_6^i$.
They are already induced at Born level.  The leading order results can
be found in \cite{Bernbra}.  We give them here in a compact form for
completeness.
\par \ \par
\underline{$q\bar{q}\to t\bar{t}$}\par Define $\displaystyle
\kappa_q=\pi^2\alpha_s^2\frac{N_C^2-1}{N_C^2}$.  Then
\begin{eqnarray}\label{qqres}
A^{q} \ &=&\ \kappa_q(2+(z^2-1)\beta^2),  \nonumber \\
c_0^{q} \ &=&\ \kappa_q(z^2-1)\beta^2,    \nonumber \\
c_4^{q} \ &=&\ 2\kappa_q,  \nonumber \\
c_5^{q}  \ &=&\ 2\kappa_q\beta^2\left(1+\frac{z^2\beta^2}
{(1+\sqrt{1-\beta^2})^2}\right),
\nonumber \\
c_6^{q} \ &=&\ -2\kappa_q\frac{z\beta^2}{1+\sqrt{1-\beta^2}}.
\end{eqnarray}
\par
\underline{$gg\to t\bar{t}$}\par Using the abbreviation $ \displaystyle
\kappa_g= \frac{\pi^2\alpha_s^2}{(1-z^2\beta^2)^2}
\frac{N_C^2-2+N_C^2z^2\beta^2}{N_C(N_C^2-1)}$ we have
\begin{eqnarray}\label{ggres}
A^g \ &=&\   2\kappa_g\left[1+2\beta^2(1-z^2)(1-\beta^2)-\beta^4z^4\right],
\nonumber \\
c_0^{g} \ &=&\ -2\kappa_g\left[(1-\beta^2)^2+\beta^4(1-z^2)^2)\right],
  \nonumber \\
c_4^{g} \ &=&\ 4\kappa_g(1-z^2)\beta^2,  \nonumber \\
c_5^{g} \ &=&\  -4\kappa_g\beta^2\left(1-2\beta^2+z^2\beta^2-
\frac{z^2\beta^4(1-z^2)}{(1+\sqrt{1-\beta^2})^2}\right),\nonumber \\
c_6^{g}  \ &=&\ -4\kappa_g\frac{z(1-z^2)\beta^4}{1+\sqrt{1-\beta^2}}. 
\end{eqnarray}
\par \ \par
Due to Bose symmetry, $R^g(\bfp,\bfk)=R^g(-\bfp,\bfk)$, which means
that $c_{0,4,5}^g$ are even functions of $z$ whereas $c_6$ is odd
in $z$. In leading order, the structure functions for $q\bar{q}\to t\bar{t}$
have the same symmetry properties, because the reaction proceeds through a
single virtual intermediate gluon, implying that also $R^q$ is invariant
under $\bfp\to -\bfp$.
\par
We may now construct a list of  ``observables'' built from
the spin operators of $t$ and $\bar{t}$:
\begin{eqnarray} \label{partobs}
\hat{\cal{O}}^i_1\ &=&\ \bfsig\otimes\bfsig,\nonumber \\
\hat{\cal{O}}^i_2\ &=&\ \hat{\bfp}\cdot\bfsig\otimes\hat{\bfp}\cdot\bfsig,
\nonumber \\
\hat{\cal{O}}^i_3\ &=&\ \hat{\bfk}\cdot\bfsig\otimes\hat{\bfk}\cdot\bfsig,
\nonumber \\  
\hat{\cal{O}}^i_4\ &=&\ (\hat{\bfp}\cdot\bfsig\otimes\hat{\bfk}\cdot\bfsig+
\hat{\bfp}\cdot\bfsig\otimes\hat{\bfk}\cdot\bfsig)/2 \ \ \ \ \ (i=q,g).
\end{eqnarray}
At the level of $t\bar{t}$ production from partons the expectation
values of these quantities for fixed $\hat{s}$ and $z$ are defined as
\begin{equation}
\langle \hat{\cal{O}}^i_{1,2,3,4}\rangle = \frac{{\mbox{tr}}(R^i 
\hat{\cal{O}}^i_{1,2,3,4})}{{\mbox{tr}} (R^i)}
,\end{equation}
where the trace is over the spin spaces of $t$ and $\bar{t}$.
They are given by linear combinations of the functions 
$c_0^i(\hat{s},z),\ldots,c_6^i(\hat{s},z)$, divided by the 
unpolarized differential cross
section $A^i(\hat{s},z)$. In particular, we find
\begin{equation}\label{partonobs1}
 \langle \hat{\cal{O}}^q_1\rangle=1,
\end{equation}
which can be easily understood from the fact that the $t\bar{t}$ pair
is produced from a single spin one boson in $q\bar{q}$ collisions at
Born level. At threshold, one can show that the quantum numbers of the
top quark pair are given by $^3S_1$ for $q\bar{q}\to t\bar{t}$ and
$^1S_0$ for $gg\to t\bar{t}$ \cite{Hara},\cite{Arens}.  In particular,
\begin{equation}
 \lim_{\beta\to 0}\langle \hat{\cal{O}}^g_1\rangle=-3,
\end{equation}
which can be verified using (\ref{ggres}). As a further example
we plot in fig.1 the rapidity distribution of  $\langle\hat{\cal{O}}^q_3\rangle$,
defined by 
\begin{equation} \label{partonrap}
\frac{\int_{-1}^{1} dz\ {\mbox{tr}}(R^q\hat{\cal{O}}_3^q)
 \delta(\hat{r}_t-\hat{r}_t')}{\int_{-1}^{1}dz\ {\mbox{tr}}(R^q)}
\end{equation}
with
\begin{equation}
\hat{r}_t=
\frac{1}{2}\ln\left(\frac{1+\beta z}{1-\beta z} \right)
\end{equation}
at a value $\beta=0.5$. We choose the rapidity of the
top quark in the partonic c.m. system as variable
here instead of the scattering angle for later comparison with
observables for $pp,p\bar{p}\to t\bar{t}X\to\ldots$ , for which
rapidity is a convenient variable.\par The expectation value $\langle
\hat{\cal{O}}^i_3\rangle$ corresponds to a helicity correlation studied by
Stelzer and Willenbrock \cite{spinreflast} and also by Mahlon and
Parke \cite{MP}, namely
\begin{equation}
\langle \hat{\cal{O}}^i_3\rangle=-\frac{d\hat{\sigma}^i(t_R\bar{t}_R
+t_L\bar{t}_L)/dz-d\hat{\sigma}^i(t_R\bar{t}_L+t_L\bar{t}_R)/dz}
{d\hat{\sigma}^i(t_R\bar{t}_R
+t_L\bar{t}_L)/dz+d\hat{\sigma}^i(t_R\bar{t}_L+t_L\bar{t}_R)/dz}.
\end{equation}
Here, the indices L and R correspond to left- and right handed
particles, respectively, and $d\hat{\sigma}^i$ denotes the partonic
cross section, $i=q,g$.  We find agreement with the values given in
\cite{spinreflast} for this correlation integrated over the scattering
angle and folded with the parton distribution functions both for
Tevatron and LHC energies ($+40$\% and $-31$\%, respectively).  Rather
than to specialize on finding ways to trace this correlation of
helicities, we will in the following try to extract as much
information on the spin density matrix as possible, i.e. construct
angular correlations among the $t\bar{t}$ decay products in close
correspondence to all four ``partonic'' observables defined in
(\ref{partobs}).
\section{Angular correlations for semileptonic $t\bar{t}$ decays}
The spin-spin correlations discussed in the previous section must be
traced in the decay products of the $t$ and $\bar{t}$.  The spin
information in the production density matrices $R^i$ is transferred to
the decay products through the parity violating decays of the top
quarks. For semileptonic decays $t\to \ell^+\nu_{\ell^+} b$, the
normalized {\it{decay spin density matrix}} $\rho$ of the top quark in
leading order (and using the narrow width approximation for the $W$
boson) reads (see, e.g., \cite{decay})
\begin{equation}\label{leptonic}
\rho(t\to \ell^+\nu_{\ell^+} b)=\frac{6x_{\ell^+}(1-x_{\ell^+})}
{(1+2\omega)(1-\omega)^2}\left[\one+\hat{\bf q}_{\ell^+}^*\cdot\bfsig\right]
\frac{dx_{\ell^+}d\Omega_{\ell^+}}{4\pi},
\end{equation}
where  $\omega=m_W^2/m_t^2$, 
$x_{\ell^+}=2E_{\ell^+}^*/m_t\in[\omega,1]$ is the scaled energy of the
lepton, and $\hat{\bf q}_{\ell^+}^*$ is the
direction of the lepton. The asterisk refers to the rest system of the
decaying quark.  For hadronic decays, $t\to W^+b\to bq\bar{q}'$, we
have, if we use the tagged $b$ quark as spin analyzer,
\begin{equation}\label{hadronic}
\rho(t\to W^+b \to b q\bar{q}')=
\left[\one+\frac{2\omega-1}{2\omega+1}\hat{\bf q}_{b}^*\cdot\bfsig\right]
\frac{d\Omega_{b}}{4\pi}.
\end{equation} 
The corresponding decay spin density matrices $\bar{\rho}$ for the
top antiquark are derived from the above ones by the replacements
$x_{\ell^+}\to x_{\ell^-},\ \hat{\bf q}_{\ell^+}^*\to -\hat{\bf
  q}_{\ell^-}^*$ and $d\Omega_{\ell^+}\to d\Omega_{\ell^-}$ in
(\ref{leptonic}) and by $\hat{\bf q}_{b}^*\to -\hat{\bf
  q}_{\bar{b}}^*,\ d\Omega_{b}\to d\Omega_{\bar{b}}$ in
(\ref{hadronic}). We may also use the $W$ boson as spin analyzer
instead of the $b$ quark. We get the corresponding decay density
matrix from (\ref{hadronic}) by 
$\hat{\bfq}^*_b\to -\hat{\bfq}_{W^+}^*,\ d\Omega_b\to d\Omega_{W^+}$.
In particular, the $W$ boson has the same spin analyzer quality as the
$b$ quark. 
The expectation value of any observable constructed
from the momenta of the final state particles will involve a trace
over the spin spaces of $t$ and $\bar{t}$ of the form ${\mbox{tr}}
\left[R^i\rho(t\to X)\otimes \bar{\rho}(\bar{t}\to \bar{X}')\right]$.
Thus the spin information of the production density matrix $R^i$ is
recovered in the decay products in a statistical sense.
\par
We will concentrate on decay channels where either the $t$ or the
$\bar{t}$ decays leptonically and the other quark decays hadronically,
i.e. on the decay modes
\begin{eqnarray}\label{semihadronic} 
t\ &\to&\ W^+ b\to  bq\bar{q}',\nonumber \\
\bar{t}\ &\to&\ \ell^{-}\bar{\nu}_{\ell^-}\bar{b}
\end{eqnarray}
and the charge conjugated ones.  These decay modes turn out to be
especially suited to construct observables which are sensitive to
spin-spin correlations: The charged lepton is the most  efficient
spin analyzer (cf.(\ref{leptonic})), while in the same event 
the momentum of the top quark
(or antiquark) may be reconstructed from its hadronic decay products.
For nonleptonic decays, since charm tagging is difficult, we will use
the bottom quark as spin analyzer. (Alternatively, we may use the $W$ boson.)
 This leads to a suppression factor
$(1-2\omega)/(1+2\omega)\approx 0.43$ in all quantities we will
discuss. One may also consider double leptonic decays where this
suppression factor is absent. However, apart from forming only $\sim
1/9$ of all $t\bar{t}$ decays, these events do not allow for a
reconstruction of the $t$ and/or $\bar{t}$ rest system on an
event-by-event basis due to the unseen neutrinos.  Any correlation
constructed from the laboratory momenta of the two charged leptons suffers 
from a large
``background'' from the unpolarized cross section, i.e. $A^i$ of eqs.
(\ref{qqres}, \ref{ggres}). In contrast, the quantities we will use get
contributions from spin-spin correlations only. 
\par In close
correspondence to the observables (\ref{partobs}) we define, for the
decay modes (\ref{semihadronic}):
\begin{eqnarray}\label{hadrobs}
 {\cal {O}}_1\ &=&\ \hat{\bfq}_b^*\cdot \hat{\bfq}_{\ell^-},\nonumber \\
 {\cal {O}}_2\ &=&\ (\hat{\bfq}_b^*\cdot\hat{\bfp}_p)
 (\hat{\bfq}_{\ell^-}\cdot\hat{\bfp}_p),\nonumber \\
 {\cal {O}}_3\ &=&\ (\hat{\bfq}_b^*\cdot\hat{\bfk}_{t})
 (\hat{\bfq}_{\ell^-}\cdot\hat{\bfk}_{t}),\nonumber \\
 {\cal {O}}_4\ &=&\ \left[(\hat{\bfq}_b^*\cdot\hat{\bfp}_p)
 (\hat{\bfq}_{\ell^-}\cdot\hat{\bfk}_{t})+
(\hat{\bfq}_b^*\cdot\hat{\bfk}_{t})
 (\hat{\bfq}_{\ell^-}\cdot\hat{\bfp}_{p})\right]/2.
\end{eqnarray}
Here, quantities without an asterisk are defined in the laboratory
frame, carets denote unit vectors and $\hat{\bfp}_p$ is the beam direction.  
The top quark
momentum $\bfk_{t}$ has to be reconstructed from its hadronic
decay products for a measurement of ${\cal{O}}_1,\ldots {\cal{O}}_4$.
The momentum of the $b$ quark was boosted into the top quark rest
system\footnote{A note of caution: The top quark rest frame defined
  in equation (\ref{leptonic}) and (\ref{hadronic}) is different from
  the one defined in (\ref{hadrobs}); the former is defined
  through a rotation-free boost from the partonic c.m. system (where
  the matrix $R^i$ is defined), while the latter is related to the
  hadronic c.m. system through a pure boost. Thus they differ by a
  Wigner rotation which has to be taken into account in the
  theoretical calculation of the expectation values of the observables
(\ref{hadrobs})
.}. 
Analogous
observables $\bar{\cal{O}}_{1,2,3,4}$ may be defined for the charge 
conjugated decay modes.
The observable ${\cal{O}}_4$ gives zero
if integrated over a symmetric rapidity interval in our leading order
calculation because, as discussed above,  
$R^{g,q}(\bfp,\bfk)=R^{g,q}(-\bfp,\bfk)$.
We therefore also define 
\begin{equation}
{\cal{O}}_5={\mbox{sign}}(r_t){\cal{O}}_4,
\end{equation}
with  
\begin{equation}
r_t=\frac{1}{2}\ln\left(\frac{E_t+\hat{\bfp}_p\cdot\bfk_t}
  {E_t-\hat{\bfp}_p\cdot\bfk_t}\right).
\end{equation}
\par
We evaluated the correlations $\langle{\cal{O}}_j\rangle \ 
(j=1,2,3,5)$ for $p\bar{p}$ collisions between $1.6$ and $4$ TeV and
for $pp$ collisions between $8$ and $16$ TeV with $m_t=180$ GeV. We
found only a weak dependence on the choice of the parton distribution
functions.  In the results below, we used the parametrization
\cite{DO} with $Q^2=4m_t^2$.  We applied cuts on the top quark
transverse momentum $|\bfk_t^T|$ and rapidity: For the lower energies, we used
$|\bfk_t^T|>15$ GeV, $|r_t|<2$, for the higher energies, the cuts
$|\bfk_t^T|>20$ GeV, $|r_t|<3$ were imposed.  
\par We will first discuss the
case of $p\bar{p}$ collisions. The results are shown in fig. 2. The
correlations are largest for small c.m. energies; there the
correlation $\langle {\cal{O}}_2\rangle$ reaches a value $\sim 3.5\%$.
The analogous correlations for the charge conjugated decays of the top
quark pair have exactly the same values due to CP invariance.  It
might seem surprising that the effects are quite small remembering
that we had spin-spin correlations of order one at the level of
$t\bar{t}$ production from partons, cf. (\ref{partonobs1}). 
The suppression comes about as
follows: As mentioned before, we lose a factor of $0.43$ by using the
$b$ quark (or $W$ boson) as spin analyzer.  
Moreover, integrating over the directions of
the $b$ quark and of the charged lepton generates roughly a factor of
$1/9$; thus the magnitude of an angular  correlation built from
these directions  is $\sim 5\%$.  This is the price we have to pay for
using observables which can be measured on an event-by-event basis and
are strictly zero in the absence of spin-spin correlations. We
will now consider signal-to-noise ratios in order to estimate
the statistical significance of the correlations. The statistical
fluctuations of our observables are given by $\Delta
{\cal{O}}_{j}=\sqrt{\langle {\cal{O}}_{j}^2 \rangle -\langle
  {\cal{O}}_{j} \rangle^2}$. We find $\Delta {\cal{O}}_1\approx
0.58$ for all energies (since $\langle {\cal{O}}_{1}^2
\rangle=1/3$) and, at $\sqrt{s}=1.8$ TeV, $\Delta {\cal{O}}_2\approx
0.36$, $\Delta {\cal{O}}_3\approx 0.36$, $\Delta {\cal{O}}_5\approx
0.25$. Measuring the analogous correlations for the charge
conjugated decays increases the statistical sensitivity. 
For example, if $N_{b\ell^-}$ denotes the number of
$b$-tagged, reconstructed events of type (\ref{semihadronic}), and we
have the same number of events in the charge conjugated channel, we
get a statistical significance $S_2$ for the combined correlations
$\langle{\cal{O}}_2\rangle$ and $\langle\bar{\cal{O}}_2\rangle$ of
\begin{equation}
S_2\equiv\frac{|\langle{\cal{O}}_2\rangle+\langle\bar{\cal{O}}_2\rangle|}
{\sqrt{2}\Delta{\cal{O}}_2}\sqrt{N_{b\ell^-}}\approx 0.14\sqrt{N_{b\ell^-}}
\end{equation}
at $\sqrt{s}=1.8$ TeV. In order to establish the spin-spin correlation
at the 3$\sigma$ level, we would therefore need $N_{b\ell^-} \approx
500$, which is in reach of the upgraded Tevatron. 
For the other three correlations shown in fig. 2, we find the
following statistical sensitivities at $\sqrt{s}=1.8$ TeV, again
combining the decay modes (\ref{semihadronic}) with the charge
conjugated ones: 
$S_1\approx 0.076\sqrt{N_{b\ell^-}}$, $S_3\approx 0.077\sqrt{N_{b\ell^-}}$, 
$S_5\approx 0.13\sqrt{N_{b\ell^-}}$.
\par 
In fig. 3  we show the rapidity distributions 
$\langle {\cal{O}}_1\delta(r_t-r_t')\rangle,\ldots,\langle 
{\cal{O}}_4\delta(r_t-r_t')\rangle$
for $\sqrt{s}=1.8$ TeV. Note the similarity of  
$\langle {\cal{O}}_3\delta(r_t-r_t')\rangle$ (dotted line in fig. 3) and the  
corresponding rapidity distribution (\ref{partonrap}) of the 
correlation $\langle {\cal{O}}_3^q\rangle$ in fig. 1. 
This similarity in form is to be expected,
since the partonic process $q\bar{q}\to t\bar{t}$ dominates at Tevatron
energies. It also holds for the other correlations and their
counterparts in $q\bar{q}\to t\bar{t}$. 
\par We now turn to $pp$ collisions. All four correlations in this
case only depend  weakly on the $pp$ c.m. energy. Between
$\sqrt{s}=8-16$ TeV they take the values
$\langle {\cal{O}}_1\rangle \approx (-2.37)-(-2.23)\%$,
$\langle {\cal{O}}_2\rangle \approx (-0.13)-(-0.24)\%$,
$\langle {\cal{O}}_3\rangle \approx (-0.72)-(-0.68)\%$, and
$\langle {\cal{O}}_5\rangle \approx (-0.25)-(-0.33)\%$.
Note especially the smallness of $\langle {\cal{O}}_2\rangle$, which
in the case of $p\bar{p}$ collisons around $\sqrt{s}\sim 2$ TeV
was the most sensitive correlation.
 We again get the same numbers for the charge conjugated decay modes. 
 At $\sqrt{s}=14$ TeV, we also determined the statistical 
fluctuations of the observables;
as mentioned before, $\Delta {\cal{O}}_1\approx 0.58$, and the other
values are
$\Delta {\cal{O}}_2\approx 0.42$, $\Delta {\cal{O}}_3\approx 0.38$,
$\Delta {\cal{O}}_5\approx 0.29$.
Assuming again an equal number of events in the decay channel 
(\ref{semihadronic}) and the charge conjugated one, we get by 
combining both correlations the following statistical sensitivities at
the LHC:
${\cal{S}}_1\approx 0.055\sqrt{N_{b\ell^-}}$,
${\cal{S}}_2\approx 0.007\sqrt{N_{b\ell^-}}$,
${\cal{S}}_3\approx 0.025\sqrt{N_{b\ell^-}}$, and
${\cal{S}}_5\approx 0.016\sqrt{N_{b\ell^-}}$.
Here it is useful to consider also
\begin{equation}
{\cal{O}}_6\equiv (\hat{\bfp}_p\times \hat{\bfq}_b^*)\cdot
                  (\hat{\bfp}_p\times \hat{\bfq}_{\ell^-})=
{\cal{O}}_1-{\cal{O}}_2.
\end{equation}
Since 
$\langle{\cal{O}}_6^2\rangle=\langle{\cal{O}}_1^2\rangle-
\langle{\cal{O}}_2^2\rangle$, 
we find for the combined correlations $\langle{\cal{O}}_6\rangle$ and
 $\langle\bar{\cal{O}}_6\rangle$ a statistical sensitivity
of ${\cal{S}}_6\approx 0.073 \sqrt{N_{b\ell^-}}$ at the LHC.
Note that in ${\cal{O}}_6$ effects of the boost from the
partonic c.m. system to the laboratory frame drop out.

The  effects are smaller at the LHC than at the 
Tevatron, but we have many more
events. For example, assuming  $N_{b\ell^-}=10^4=N_{\bar{b}\ell^+}$, 
we can establish the spin-spin correlations to $7.3\sigma$ by measuring 
${\cal{O}}_6$ and $\bar{\cal{O}}_6$.
In fig. 4 we show the rapidity distributions of the correlations
at $\sqrt{s}=14$ TeV.
Due to the dominance of gluon fusion at LHC energies, the shapes
of the curves are similar to the corresponding rapidity distributions
of the correlations at parton level $\langle{\cal{O}}_{1,2,3,4}^g\rangle$ 
defined in (\ref{partobs}).

\section{Conclusions}
We have shown that the sizable spin-spin correlations induced
at leading order QCD in $q\bar{q}\to t\bar{t}$ and $gg\to t\bar{t}$
may be measured both at the Tevatron and the LHC. For such
measurements,
 angular correlations among the decay products in ``semihadronic''
$t\bar{t}$ decays are especially suited. 
The correlations we propose get nonzero 
contributions only from $t\bar{t}$ spin-spin correlations --- which are
predicted by perturbative QCD.  
All calculations were performed at leading order and using the
narrow width approximation for the top quark.
At present, NLO corrections are only known for the functions $A^{q,g}$ in
(\ref{qqres}), (\ref{ggres})
which determine the production rate of $t\bar{t}$ pairs \cite{BKNS}.
For a more refined theoretical study of the spin-spin correlations,
we would need the complete spin density matrix in NLO as well as
an estimate of the effects of non-factorizable contributions
\cite{Melni}. In summary, the experimental study of spin-spin
correlations of top quark pairs seems very interesting and feasible.
\par
\bigskip
\noindent {\bf Acknowledgements}\par\noindent
I would like to thank W. Bernreuther, L. M. Sehgal and P. Uwer for
helpful discussions, and the CERN TH Division, where part of
this work was done, for hospitality and partial support.
\bigskip
\vfil\eject

\vfil\eject
{\Large{\bf Figure Captions}}
\medskip
\begin{description}
\item [Fig. 1] Rapidity distribution (\ref{partonrap}) of 
$\langle \hat{\cal{O}}^q_3\rangle$ at $\beta=0.5$.
\item [Fig. 2] Correlations $\langle {\cal{O}}_1\rangle$ (full line),
  $\langle {\cal{O}}_2\rangle$ (dashed line),
 $\langle {\cal{O}}_3\rangle$ (dotted line), and 
 $\langle {\cal{O}}_5\rangle$ (dash-dotted line)
as a function of the c.m. energy for $p\bar{p}$ collisions.
\item [Fig. 3] Rapidity distributions 
$\langle {\cal{O}}_1\delta(r_t-r_t')\rangle$ (full line),
$\langle {\cal{O}}_2\delta(r_t-r_t')\rangle$ (dashed line),
$\langle {\cal{O}}_3\delta(r_t-r_t')\rangle$ (dotted line), and
$\langle {\cal{O}}_4\delta(r_t-r_t')\rangle$ (dash-dotted line),
at $\sqrt{s}=1.8$ TeV for $p\bar{p}$ collisions.
\item [Fig. 4] Same as Fig. 3, but for $pp$ collisions
at $\sqrt{s}=14$ TeV.
\end{description}
\vfil\eject
\setlength{\unitlength}{1cm}
\begin{picture}(15,10)
\hskip 2.5cm 
\epsfysize=12cm
\epsfbox{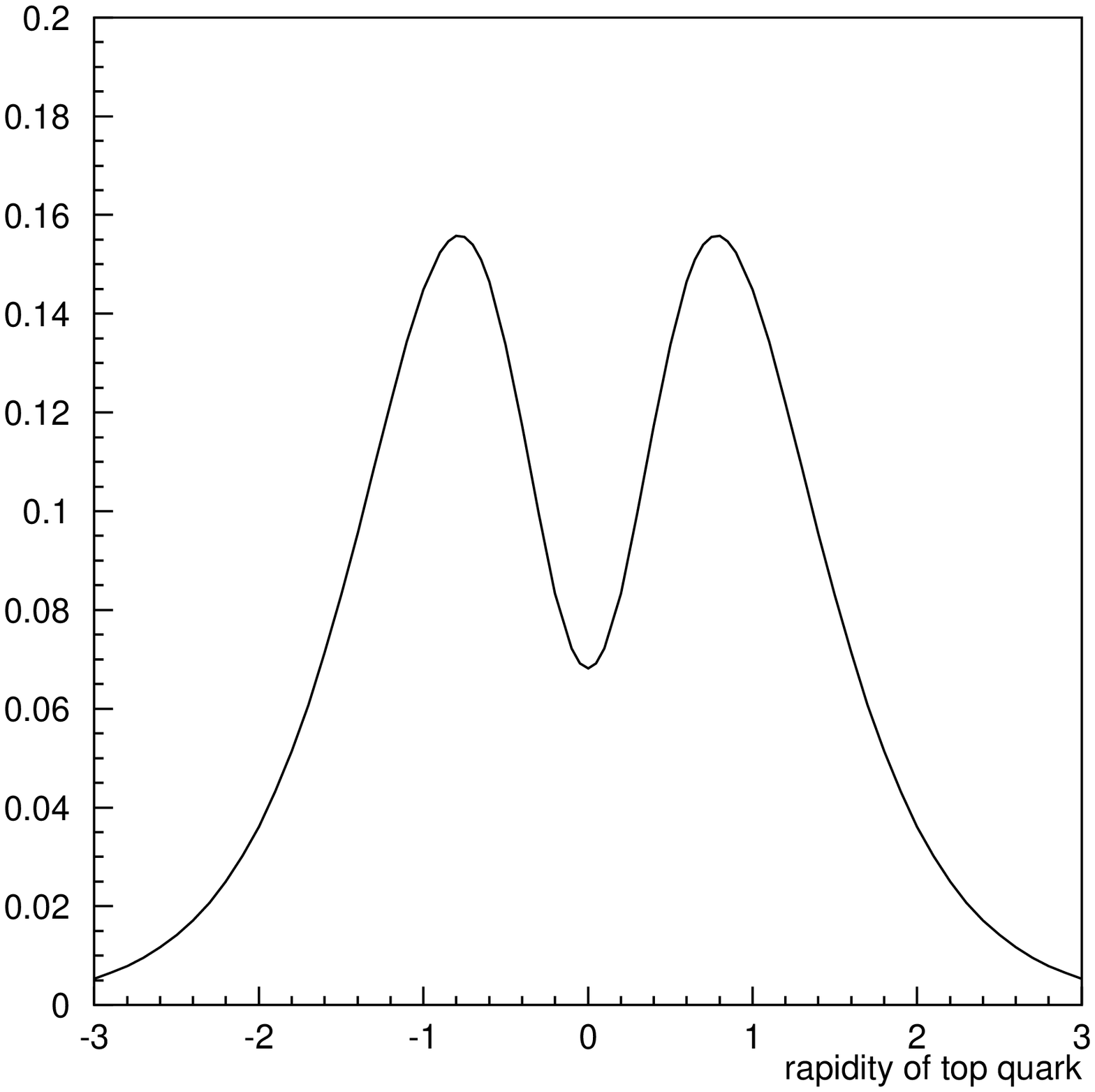}
\end{picture}
\ 
\vskip-2.75cm
\ 
\begin{center}
{\bf Fig. 1}
\end{center}
\ 
\vskip 0.75cm
\ 
\begin{picture}(15,10)
\hskip 2.5cm 
\epsfysize=12cm
\epsfbox{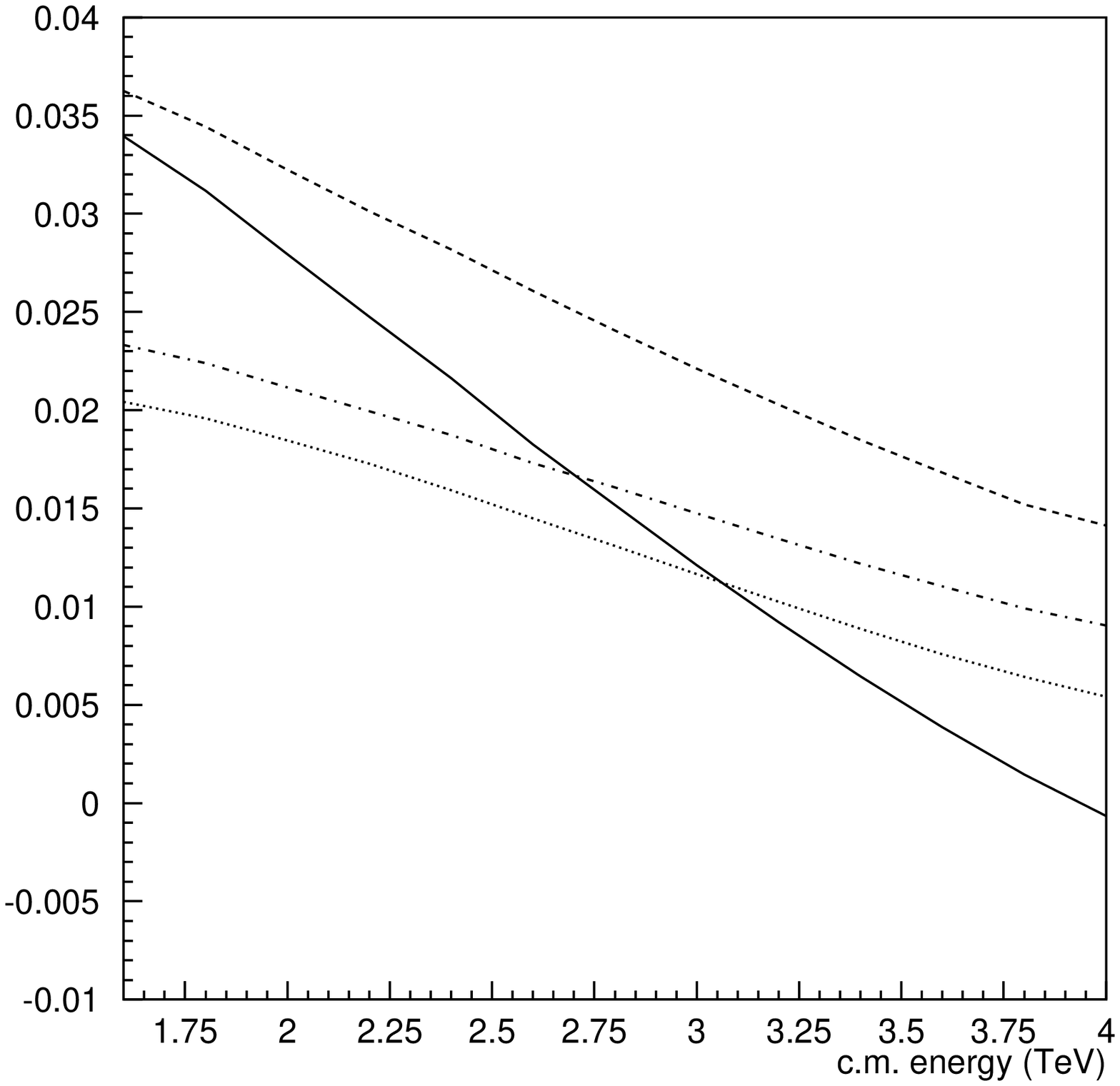}
\end{picture}
\ 
\vskip -2.75cm
\ 
\begin{center}
{\bf Fig. 2}
\end{center}
\
\vskip 0.75cm
\ 
\begin{picture}(15,10)
\hskip 2.5cm 
\epsfysize=12cm
\epsfbox{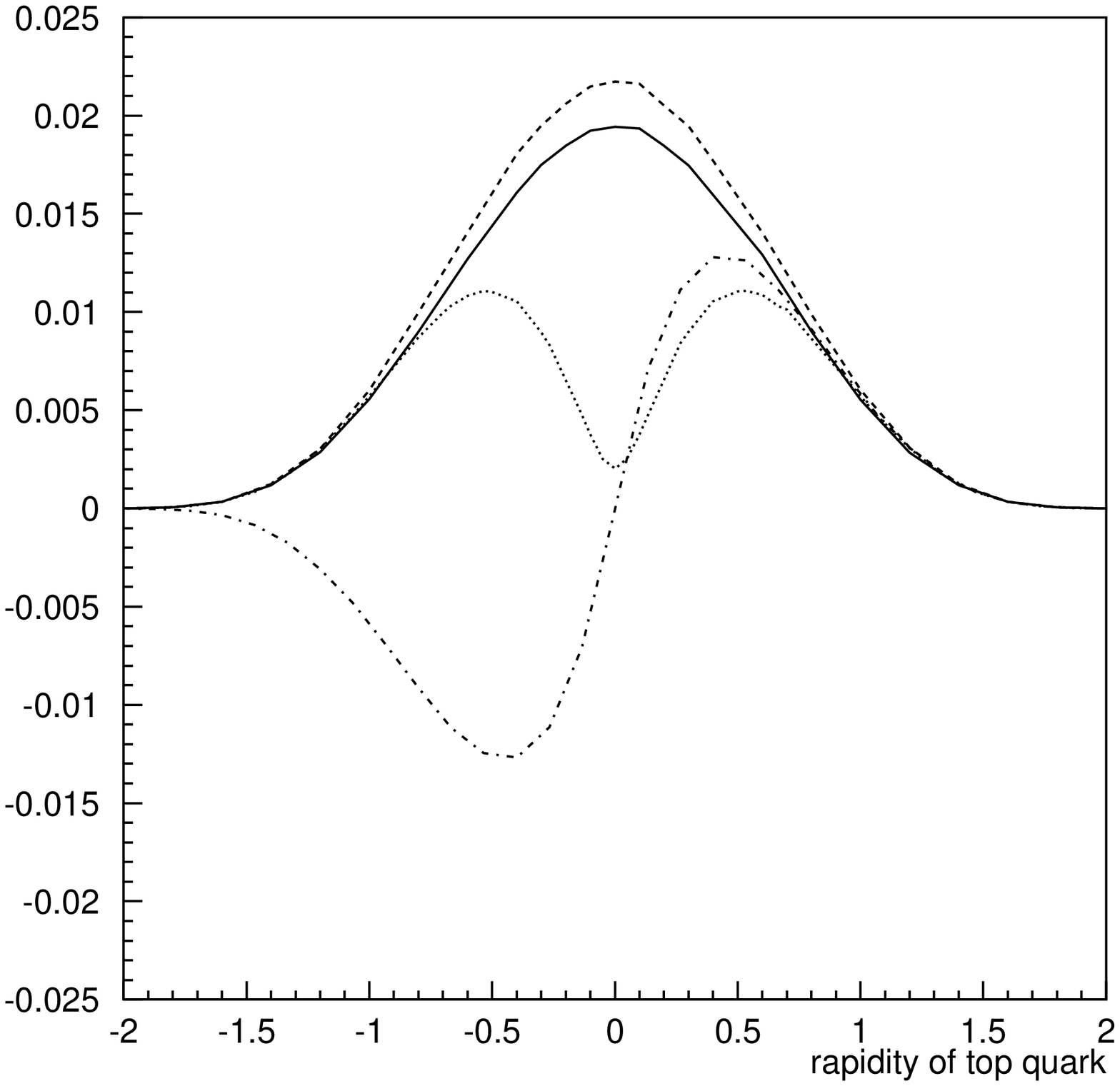}
\end{picture}
\ 
\vskip -2.75cm
\ 
\begin{center}
{\bf Fig. 3}
\end{center}
\ 
\vskip 0.75cm
\ 
\begin{picture}(15,10)
\hskip 2.5cm 
\epsfysize=12cm
\epsfbox{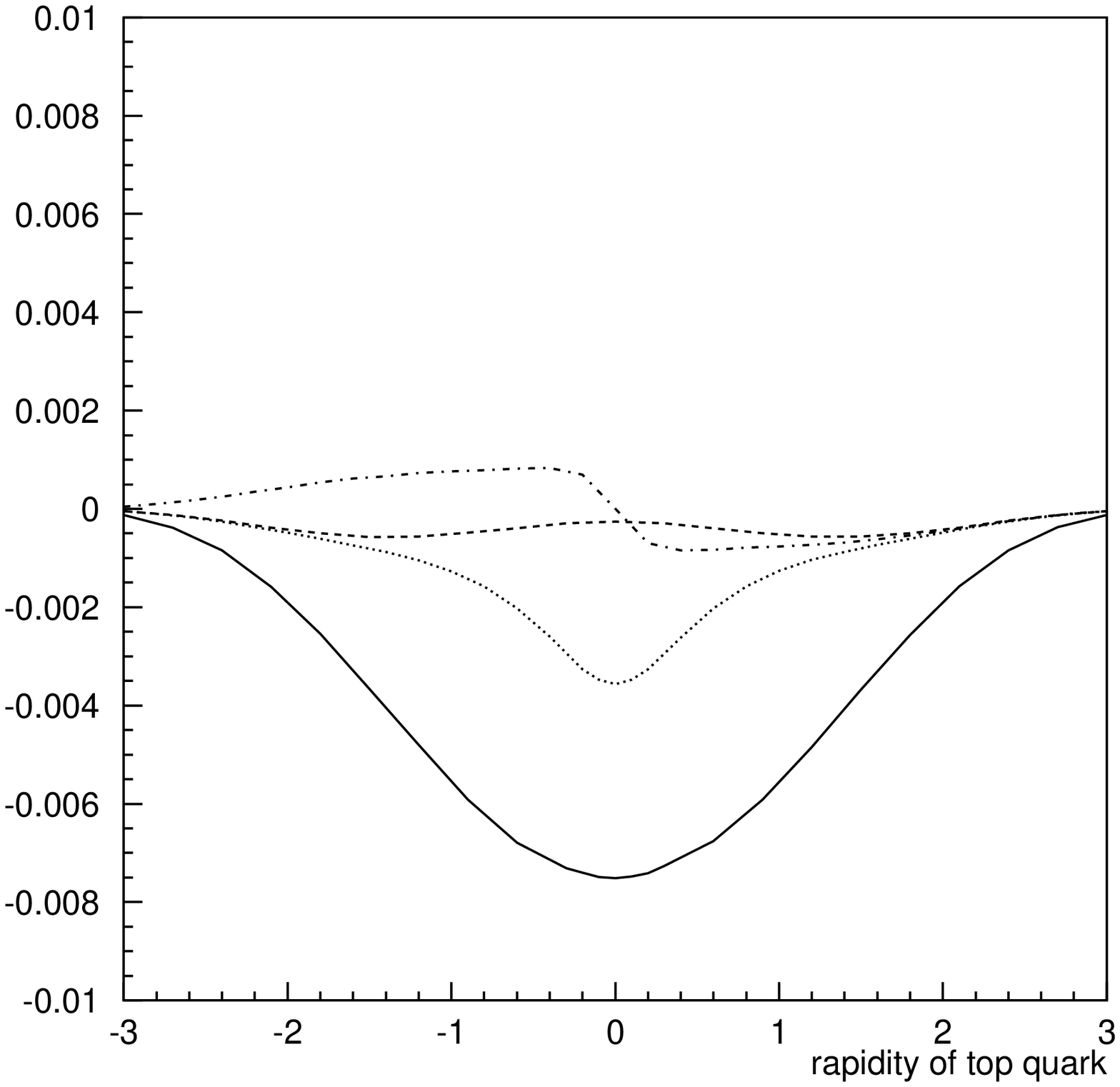}
\end{picture}
\ 
\vskip -2.75cm
\  
\begin{center}
{\bf Fig. 4}
\end{center}

\end{document}